\documentclass[bibyear]{aa} 

\usepackage{graphicx}
\usepackage{txfonts}
\usepackage{amsmath}
\usepackage{latexsym}
\usepackage{amsfonts}
\usepackage[normalem]{ulem}
\usepackage{soul}
\usepackage{array}
\usepackage{amssymb}
\usepackage{extarrows}
\usepackage{graphicx}
\usepackage{textcomp}
\usepackage{subfig}
\usepackage{wrapfig}
\usepackage{txfonts}
\usepackage{wasysym}
\usepackage{enumitem}
\usepackage{adjustbox}
\usepackage{ragged2e}
\usepackage[svgnames,table]{xcolor}
\usepackage{tikz}
\usepackage{longtable}
\usepackage{changepage}
\usepackage{setspace}
\usepackage{hhline}
\usepackage{multicol}
\usepackage{tabto}
\usepackage{float}
\usepackage{multirow}
\usepackage{makecell}
\usepackage{fancyhdr}
\usepackage[toc,page]{appendix}
\usepackage[hidelinks]{hyperref}
\usetikzlibrary{shapes.symbols,shapes.geometric,shadows,arrows.meta}
\tikzset{>={Latex[width=1.5mm,length=2mm]}}
\usepackage{flowchart}
\usepackage{float}

\begin{document}

\title{Recent Observations of the Rotation of Distant Galaxies and the Implication for Dark Matter}
\titlerunning {Rotation of Distant Galaxies and the Implication for Dark Matter}
\author{A. H. Nelson, P. R. Williams }
\author{A. H. Nelson, \& P. R. Williams }
\institute{School of Physics and Astronomy, Cardiff University, Cardiff CF24 3AA, United Kingdom\\\email{nelsona@cf.ac.uk}}
\date{Received December 4, 2023}

\abstract
{Recent measurements of gas velocity in the outer parts of high redshift galaxies suggest that steeply falling rotation curves may be common, or even universal, in these galaxies, in contrast to the near universal flat, non-declining rotation curves in nearby galaxies.}
{To investigate the implications of these postulated steeply falling rotation curves for the role of dark matter in galaxy formation.}
{Using an established computer code, the collapse of dark matter and baryonic matter together, starting with a variety of initial conditions, is simulated for comparison with the observed rotation curves.}
{As soon as a smooth stellar disc is formed in the baryonic matter, with properties similar to the observed high redshift galaxies, the computed rotation curves are, without exception, relatively flat to large radius in the gas disc. Only a simulation without a dark matter halo is able to reproduce the observed rotation curves.}
{This would imply that, if the high redshift steeply falling rotation curves turn out to be common, then the standard scenario for galaxy formation for these galaxies, namely baryonic matter falling into the potential well of a massive dark matter halo, must be wrong, unless there is pressure support via velocity dispersion significantly higher than has so far been observed. It would also imply that for these galaxies the flat rotation curves at low redshift must be due to dark matter which has subsequently fallen into the galactic potential well, or there must be some other explanation for their contemporary flat rotation curves, other than dark matter.}


\maketitle

\section{Introduction}

\vspace{\baselineskip}

It is well known that the rotation curves of nearby spiral galaxies, as observed in the emission lines of the gaseous disc, exhibit high velocities compared to what would be expected to be supportable in the combined gravitational field of the gaseous and stellar components of the galaxy. This is interpreted as evidence of a massive extended dark matter halo, coincident with the visible matter, resulting in the flat rotation curve out to large radii beyond the effective stellar radius (Rubin \& Ford 1970; Rubin, Ford, \& Thonnard 1978; Bosma 1981; Sofue \& Rubin 2001; de Blok et al. 2008). And flat rotation curves are ubiquitous in virtually all local spiral galaxies.

\begin{figure}[H]
\resizebox{\hsize}{!}{\includegraphics[width=3.5in,height=3.8in]{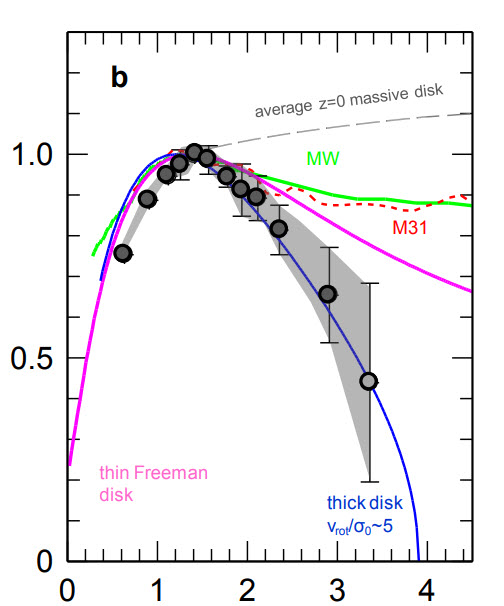}}
\caption{Rotation curve, $V/V_{max}$ versus $R/R_{1/2}$, obtained from a stack of distant galaxies (black dots with error bars) compared to the rotation curve of M31 (red broken line) and the Milky Way (green line). (Fig. 2b of Genzel et al. 2017)}
\label{fig:rot_genzel}
\end{figure}

By local we mean within a few million light years of the Milky Way, i.e with a redshift $\lesssim 0.01$. However recently sensitive enough instrumentation has been developed to obtain the rotation curves of much more distant galaxies, with redshifts $\gtrsim 1.0$. Because of their distance it is very difficult to obtain the necessary spectra to measure doppler shifts, particularly at the greatest radii outside the stellar discs, but a technique can be used to add the data from multiple galaxies in a stack (after adjustment for angular size, maximum velocity, and orientation), thus increasing the signal to noise in the outer parts of the collection of galaxies. Genzel et al. (2017) and Lang et al. (2017) have recently done this for a stack of $\sim 100$ galaxies more distant than 5 billion light years ($0.6 <  z < 2.6$), and obtained the common rotation curve shown in Fig. \ref{fig:rot_genzel} (Fig. 2b of Genzel et al. 2017)

\begin{figure}[H]
\resizebox{\hsize}{!}{\includegraphics{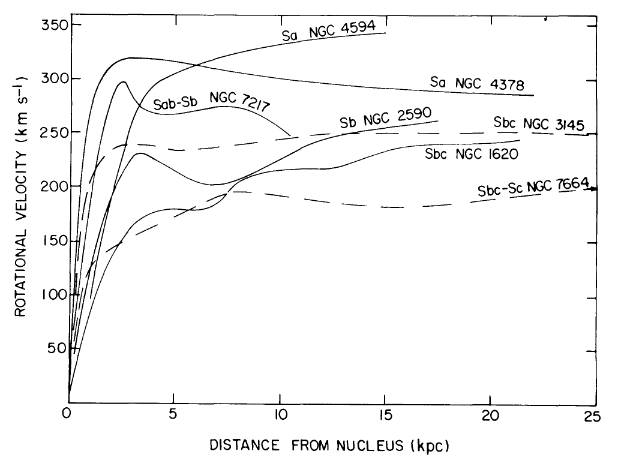}}
\caption{Rotation curves of a sample of nearby galaxies (Fig. 3 Rubin et al. 1978).}
\label{fig:rot_local}
\end{figure}

The rotation curve falls off rapidly with radius, unlike the rotation curves of nearby galaxies shown in Fig. \ref{fig:rot_local}. 

The normalisation with respect to size and maximum velocity required to stack the multiple galaxies has been criticised by Tiley (2019), who contends that normalising by the disc radius ($R_{d}$) gives different extended rotation curves from normalising by the turnover velocity radius ($R_{max}$), and in fact removes the falling velocities beyond $R_{max}$.   However Lang (2018) claims that the fall-off of the rotation curve is independent of the nomalisation method, and normalising by $R_{1/2}$ yields an even more steeply falling rotation curve beyond $R_{max}$.

Also Sharma et al. (2021) have analysed the spectra of 344 galaxy sources in the range $0.57 < z < 1.04$ and claim, after correction for kinematic modelling, beam smearing and pressure support from gaseous velocity dispersion, that 256 of their sources have individual flat rotation curves out to a radius of up to 18 kpc, where the median $R_{d}$ for the sample is $\sim 3$ kpc. One caveat which may serve to differentiate these results from those of Genzel and Lang is that Sharma's sources have a median z of 0.87, while those of Genzel and Lang have a median z of 1.52, with only a little overlap between the two sets. The lookback time difference between these two median redshifts is almost 2 Gyr which may be significant -- see the conclusion section.
 
This paper will present computer simulations of galaxy formation which show that, irrespective of the initial conditions, if a galaxy forms within a massive dark matter halo then as soon as a stellar disc is formed the gaseous disc will have a non-declining rotation curve to large radius, unlike Fig. \ref{fig:rot_genzel}. 
So if it turns out that the rotation curves for all distant galaxies are steeply falling, then this implies that galaxies form initially without a massive dark matter halo.  But this is contrary to the accepted paradigm for galaxy formation, where the dark matter halo collapses first, followed by the infall of the baryonic matter into the gravitational potential well of the dark matter, where it forms visible stars.

\section{Simulations and Computational Method}

The code used to perform the simulations is described in Williams \& Nelson (2001), Churches, Nelson \& Edmunds (2001, 2004), and Smith, Davies \& Nelson (2008, 2010). The code is a standard SPH/Treecode (Monaghan 1992; Barnes \& Hut 1986), employing a $1/(r+\epsilon)$ potential, individual particle timesteps, fully dynamic kernel radius, and star formation via a Schmidt law. The original code assumed an isothermal equation of state, on the basis that any heating of the interstellar gas would be instantly radiated away. However since the velocity dispersion in the interstellar medium has a potentially significant effect on the observed rotation curves, the code has been extended to include a full energy equation with cooling and supernova feedback to a turbulent energy component in the dynamic pressure.

The star formation rate is given by
\begin{equation}
\frac{d\rho_{\star}}{dt} = k \rho^{n},
\label{eq:schmidt}
\end{equation}
where $\rho$ is the gas density, $\rho_{\star}$ is the star mass density, and $k$ and $n$ are constants. The value 1.25 is used for $n$, within the range suggested by Kennicutt (1998) based on the gas surface density. If one makes the assumption that a gas disc with a mass of $5 \times 10^{10} M_{\odot}$, a radius of 10 kpc, and a height of 200 pc turns 90\% of its mass into stars over a timescale of 5 Gyr with $n = 1.25$, then the value of $k$ thus obtained is
\begin{equation}
k = 0.96 \  M_{\odot}^{-0.25} \text{pc}^{0.75} \text{Gyr}^{-1},
\end{equation}
which is the value used.

The reservoirs of energy in the simulation, apart from the potential and kinetic energy of the dark matter, gas, and eventually star particles, consist of $u$, the gas thermal specific energy, and $q$, the specific energy in turbulent motion created by SN feedback (Springel 2000). These two are governed by the equations
\begin{equation}
\frac{du}{dt} = - (\gamma - 1)u\nabla.\mathbf{v - \ }\Lambda_{r}\rho f(T) - \Lambda_{B}\rho u^{0.5} - \Lambda_{C}(1 + z)u + \Lambda_{t} q^{1.5}  
\end{equation}
and
\begin{equation}
\frac{dq}{dt}\  = - (\gamma - 1)q\nabla.\mathbf{v} + \Lambda_{SN}{\dot{\rho}}_{\star}\rho^{-1} - \Lambda_{t}q^{1.5}
\end{equation}
Here,
\begin{itemize}
  \renewcommand{\labelitemi}{}
  \item $\textbf{v}$ is the gas velocity,
  \item $\Lambda_{r}$ is a radiative cooling coefficient $= 1.73 \times 10^{20}$ m$^{5}$s$^{-3}$kg$^{-1}$,
  \item $\Lambda_{B}$ is a thermal Bremsstrahlung cooling coefficient $= 2 \times 10^{11}$ m$^{4}$s$^{-2}$kg$^{-1}$,
  \item $\Lambda_{C}$ is a Compton cooling coefficient $= 1.34 \times 10^{-20}$ s$^{-1}$,
  \item $\Lambda_{t}$ is a turbulent decay coefficient $= 1.7 \times 10^{-18}$ m$^{-1}$,
  \item $z$ is the redshift,
  \item $\Lambda_{SN}$ is a supernova heating coefficient $= 1.74 \times 10^{12}$ m$^{2}$ s$^{-2}$,
  \item ${\dot{\rho}}_{\star}$ is the star mass formation rate of equation \ref{eq:schmidt},
  \item $\gamma$ is the ratio of specific heats,
  \item $T$ is the temperature $= mu/3K$ where $m$ is a mean atomic mass (equal to 1.4 times the hydrogen atomic mass),
  \item $K$ is Boltzmann's constant, 
\end{itemize}
and
\begin{equation}
f(T) = 10^{{- 0.1 - 1.88(5.23 - \log T)}^{4}} + \ 10^{{- 1.7 - 0.2(6.2 - \log T)}^{4}}.
\end{equation}
The radiative and Compton cooling formulae are taken from Katz \& Gunn (1991), thermal bremmstrahlung (assuming near full ionization) from Boyd \& Sanderson (1969), and the SN feedback formula from Dalla Vecchia \& Schaye (2012).

The SN feedback method used follows the paradigm outlined by Springel (2000), except that a different formula is used for the decay of turbulent energy to thermal energy, based on the physics of the Kolmogorov cascade, that is $dq/dt \propto -q^{1.5}$.

In contrast to the cosmological scale simulations (Schaye et al. 2015; Teklu et al. 2015; Kaviraj et al. 2017; Tremmel et al. 2017; Springel 2018; Dave et al. 2019; Dubois et al. 2021; Bennett \& Sijacki 2022), which model large volumes of the early universe leading to multiple galaxy formations, we follow the approach of Katz (1992), and Steinmetz \& Muller (1995) which is to model the collapse of isolated spherical perturbations. Although this does not incorporate the interaction of the protogalaxy with nearby intergalactic gas, using $\sim 100,000$ SPH and dark matter particles allows comparable resolution to that of the individual galaxies formed in the cosmological scale simulations. The results presented here will focus on the early formation of the gas and stellar discs, and their dynamical properties, which is not greatly affected by interactions with the intergalactic environment. 

The purpose of the simulations is to investigate with various initial conditions whether the falling rotation curves reported by Genzel et al. can be replicated within the standard paradigm of a dominant dark matter component. The approach adopted gives more control over the initial conditions of the formation of an individual galaxy than is available in larger scale simulations, while we maintain a strong connection with these simulations by keeping total angular momentum and mass within the expected bounds, and this is reflected in the fact that the resulting dark matter haloes achieve a typical Navarro, Frenk \& White (NFW) density profile (see Appendix \ref{app:nfw}). 

Various configurations of the initial gas and dark matter spheres have been employed, with gas and dark particles set down in uniform density spheres, with positions displaced by small random displacements to emulate Poisson noise. The common feature is that the dark matter component has 83\% of the mass, and the baryonic star forming gas eventually falls into the final dark matter potential well. All of the initial scenarios lead to galaxies of similar structure and dynamics, the main reason being that the collapsing protogalaxies go through a violent relaxation phase during which the memory of the detailed initial structure gets erased. The amplitude and form of the initial noise in the particle positions also has little effect on the end result, a fact coroborated by Katz \& Gunn (1991). Churches (1999) and Churches, Nelson \& Edmunds (2001) have also used a range of initial conditions, and shown that the parameters which have the most significant effect on the final galaxy properties are the total mass of the system, and its angular momentum.

The effect of tidal torques by neighbouring gas and dark halos is approximated by setting the initial spheres in solid body rotation. The velocity of rotation is characterised by the dimensionless spin
parameter $\lambda$ which is given by
\begin{equation}
\lambda = J G^{- 1}{|E|}^{0.5}M^{-2.5}
\end{equation}
where $J$ is the total angular momentum, $E$ is the total energy, $M$ is the total mass of the system, and $G$ is the gravitational constant. The theoretical prediction of Peebles (1971) is that $\lambda$ is of the order of 0.1, a value that has been confirmed by N body experiments. The initial redshift is set at 100, and the initial ensemble is assumed to have stopped expanding, and starts with no radial velocity.

Five different configurations of initial conditions have been explored
\begin{itemize}
  \renewcommand{\labelitemi}{}
  \item Type 1: initial precise coincidence of the gas and dark matter spheres,
  \item Type 2: the dark matter sphere more collapsed than the gas sphere,
  \item Type 3: the initial gas and dark matter spheres not coincident,
  \item Type 4: multiple dark matter spheres embedded in a larger gas sphere,
  \item Type 5: a gas sphere as in type 1, but no dark matter.
\end{itemize}
In types 1, 3, and 4 the total mass of the system is $5 \times 10^{11}$ M$_{\odot}$, while in type 2 it is $10^{12}$ M$_{\odot}$. In all these types 83\% of the initial mass is in dark matter, and 17\% in gas. In type 5 the total mass is $8.5 \times 10^{10}$ M$_{\odot}$.  The initial radius of the gas sphere in each case is 175 kpc, and in types 2 and 3 a smaller initial radius is used for the dark matter spheres, and a rotational velocity of 0.161 Gyr$^{-1}$ is applied about the z-axis. This yields in type 1 a total angular momentum of $J = 6 \times 10^{67}$ kg m$^{2}$s$^{-1}$ and a total energy of $E = -2 \times 10^{52}$ J, and thus a spin parameter of 0.13. 

The evolution of all of these is followed to the point at which a stable stellar disc is formed with no close mergers, a stellar mass comparable to the gas disc and in the range $2-5 \times 10^{10}$ M$_{\odot}$, and with half light radius ($R_{1/2}$) in or near the range 4-9 kpc (as measured by the stellar number density modified by dust obscuration – see Appendix \ref{app:obsc}). This is the range of $R_{1/2}$ in the set of galaxies in Fig. \ref{fig:rot_genzel}.   

\section{Results}

The evolution of each type follows a similar scenario in that the baryonic gas falls into the potential well (or wells) of the collapsing dark matter. This is accompanied by violent relaxation involving strong tidal interactions disrupting and merging sub-galactic clumps, leading to a final relatively smooth and long lived spiral galaxy. The properties of the galaxies at the point where a smooth stellar disc with no merger interactions is established are summarised in tables \ref{table0} and \ref{table1}. The quantity $\sigma$ is the total velocity dispersion, incorporating $u$, $q$, and the random motion of the SPH particles relative to the bulk motion; and $\langle \sigma \rangle$ is the average over radius. Images of the gas, stars and dark matter at this point are displayed in Appendix \ref{app:images}, and there are also links in the figure captions to video animations for each of the simulations. In spite of the very different initial configurations of baryonic and dark matter, the violent relaxation intrinsic to the formation process leads to galaxies which are far from being dissimilar. The dark haloes in types 1-4 also relax to an approximate NFW profile by the time the stellar disc is formed (see Appendix \ref{app:nfw}).

\begin{table*}
\caption{Galaxy star formation properties at the point where the rotation curves in figures \ref{fig:type1_rot} to \ref{fig:type5_rot} were evaluated.}
\label{table0}
\centering
\begin{tabular}{l l l l l l l}
\hline\hline
\parbox[t]{2cm}{Type} & \parbox[t]{3cm}{$z$ at stable\\disc formation} & \parbox[t]{3cm}{Stellar mass in\\disc / $10^{10}$ M$_{\odot}$} & \parbox[t]{2cm}{Gas mass\\/ $10^{10}$ M$_{\odot}$} & \parbox[t]{3cm}{Star formation\\rate / M$_{\odot}$yr$^{-1}$} & \parbox[t]{3cm}{Maximum star\\formation\\rate / M$_{\odot}$yr$^{-1}$} \\
\hline
 1 & 0.45 & 5.2 & 1.7 & 3.7 & 15.4 \\
 2 & 1.06 & 4.5 & 3.3 & 8.3 & 20.3 \\ 
 3 & 0.81 & 2.6 & 8.5 & 4.5 & 7.5 \\
 4 & 0.89 & 4.3 & 3.4 & 4.1 & 16.6 \\
 5 & 0.78 & 2.0 & 5.9 & 140.7 & 160.8 \\
\hline
\end{tabular}
\end{table*}

\begin{table*}
\caption{Galaxy rotation curve properties at the point where the rotation curves in figures \ref{fig:type1_rot} to \ref{fig:type5_rot} were evaluated.}
\label{table1}
\centering
\begin{tabular}{c c c c c c c c c}
\hline\hline
Type & Range of $R_{1/2}$ / kpc & $V_{max}$ / km s$^{-1}$ & $\langle \sigma \rangle$ outside $R_{max}$ / km s$^{-1}$ \\
\hline
 1 & 4.2-11.8 & 211 & 30 \\
 2 & 3.7-10.6 & 225 & 37 \\
 3 & 3.1-12.4 & 246 & 42 \\
 4 & 2.6-7.5 & 294 & 44 \\
 5 & 3.7-16.8 & 195 & 59 (inside 15 kpc) \\
\hline
\end{tabular}
\end{table*}

\subsection{Type 1}

\begin{figure}[hb!]
\resizebox{\hsize}{!}{\includegraphics{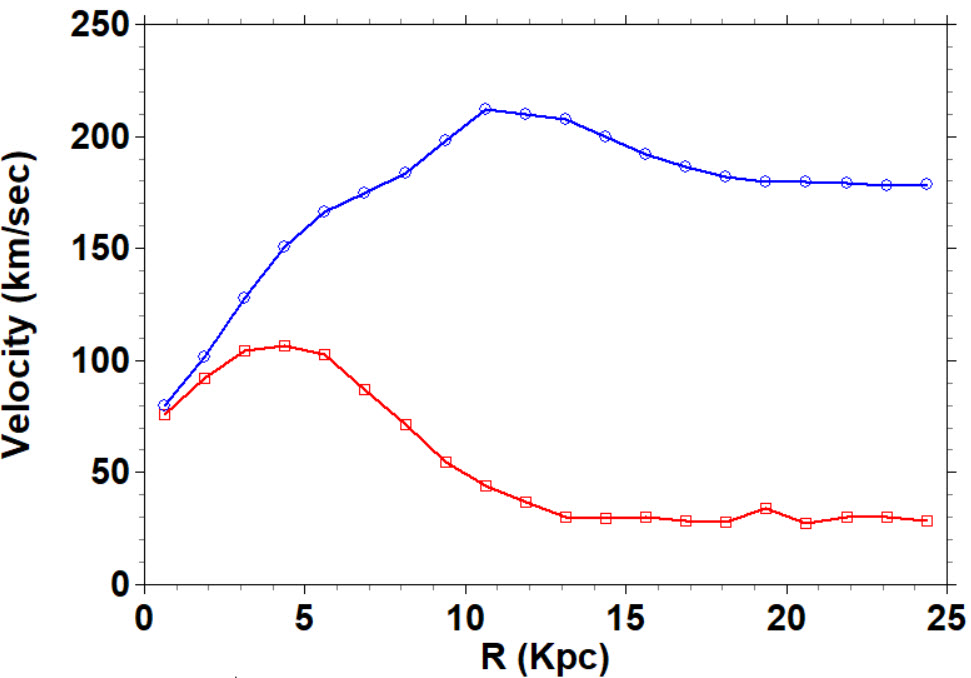}}
\caption{Type 1 rotation curve - rotational velocity (blue) and velocity dispersion (red).}
\label{fig:type1_rot}
\end{figure}

This type follows a similar scenario to the phases described in Williams \& Nelson (2001), that is there is (i) initial collapse to a high density with compression along the z-axis onto the x-y plane, then (ii) re-expansion mainly in the x-y plane, along with fragmentation into massive clumps, (iii) re-collapse accompanied by strong tidal interactions disrupting the sub-galactic clumps, leading to (iv) a long lived spiral galaxy.

A galaxy disc is formed, with $5.2 \times 10^{10}$ M$_{\odot}$ in the stellar disc and $1.75 \times 10^{10}$ M$_{\odot}$ in the gas component, 7.86 Gyrs after the start of the simulation. If we take the start of the simulation to have been at a redshift of 100, then this galaxy has been formed at $z = 0.45$. The value of $R_{1/2}$ is estimated to be between 4.2 and 11.8 kpc, while the radius of the gas disc is estimated to extend out to approximetly 70 kpc (see \ref{fig:type1} in Appendix \ref{app:images}).
  
The reason for the wide range in the estimate of $R_{1/2}$ is because we have taken the obscuration by dust generated during star formation into account, and while we are able to estimate the optical depth of the dust using Churches, Nelson \& Edmunds (2001, 2004), the distribution of gas/dust and stars in the vertical direction is not resolved by the simulations. To calculate the amount of obscuration as a function of radius and hence $R_{1/2}$, we therefore apply the two extreme cases, (1) dust concentrated in a central layer much smaller then the stellar disc thickness, and (2) dust distributed throughout the stellar disc thickness (see Appendix \ref{app:obsc}). This procedure yields values of $R_{1/2}$ equal to 4.2 and 11.8 kpc respectively in the case of type 1, and is applied to all 5 types.

The rotational velocity has a maximum of 211 km s$^{-1}$. Also in Fig. \ref{fig:type1_rot} the velocity dispersion of the gas is plotted as a function of radius. This is a combination of thermal and turbulent velocities derived from the energy equation with input from SN, together with the intrinsic velocity dispersion of the SPH particles arising from the violent relaxation during the galaxy formation process. At this time the average velocity dispersion outside $R_{max}$ is 30 km s$^{-1}$, which is insufficient to add significant support against the dark matter potential well, hence the high rotational velocity.

There is a link in the figure captions in Appendix \ref{app:images} to a video file of all the simulations showing the evolution of the gas, star and dark matter components face-on and edge-on. In the case of type 1 the main stellar disc forms at $z = 0.97$, but the video in this case runs on to $z=0.45$, since there is a significant close satellite of the main galaxy at $z = 0.97$. The satellite has merged with the main galaxy by $z = 0.45$, at which point we display the rotation curve. In all 5 simulations in this paper we concentrate on the point at which a stable stellar disc has been formed ,with no significant mergers, though the simulations all run to much lower redshift with little change in their rotation curves.

\subsection{Type 2}

\begin{figure*}[ht]
\begin{minipage}{.5\textwidth}
  \raggedright
  \includegraphics[width=\linewidth]{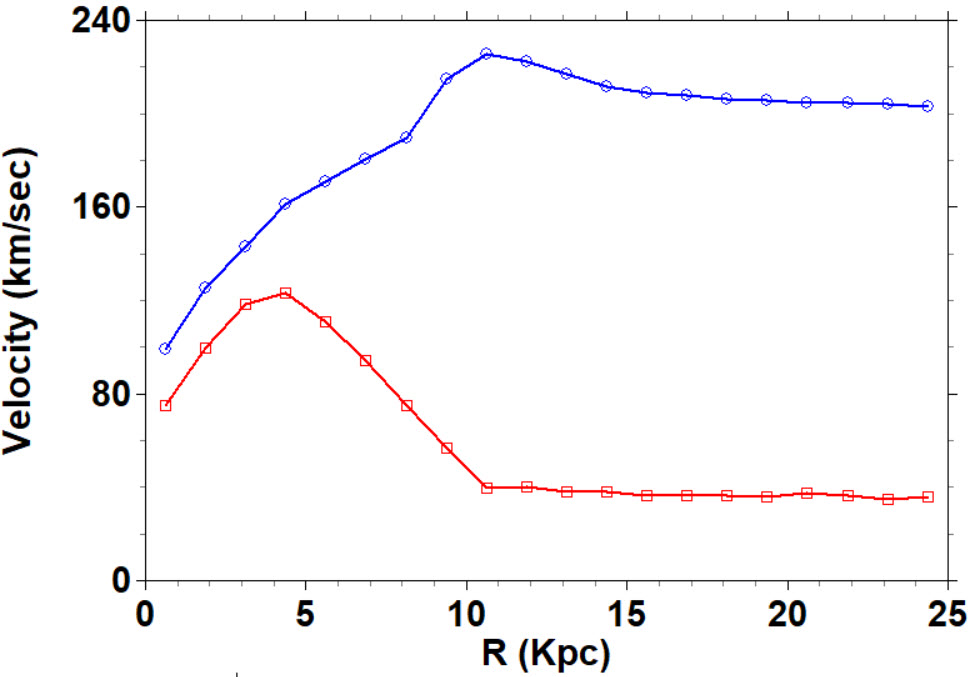}
  \caption{Type 2 rotation curve - rotational velocity (blue) and velocity dispersion (red).}
  \label{fig:type2_rot}
\end{minipage}
\begin{minipage}{.5\textwidth}
  \raggedleft
  \includegraphics[width=\linewidth]{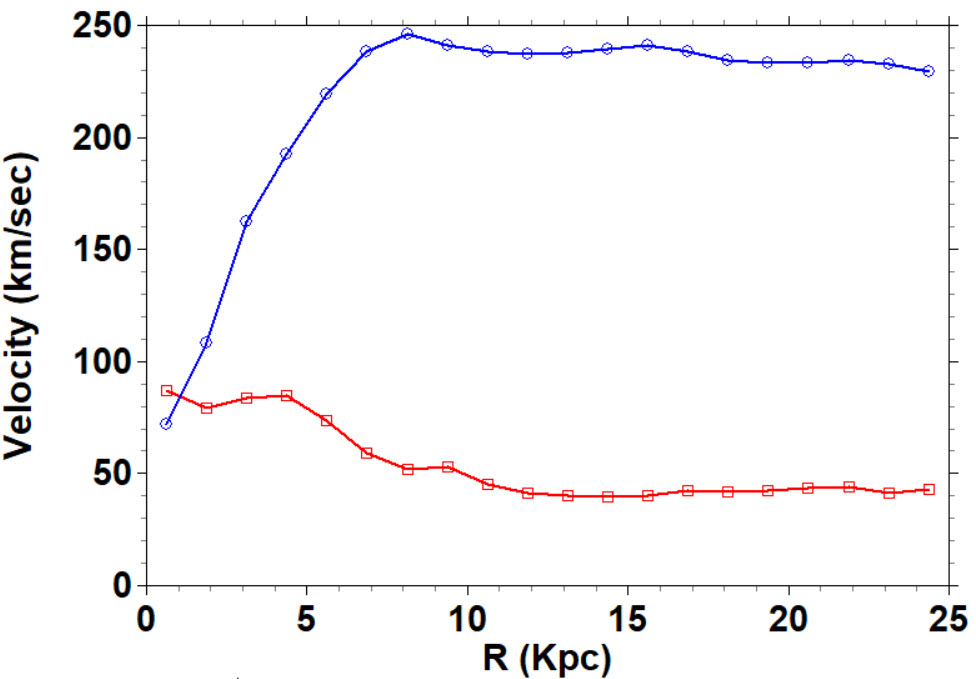}
  \caption{Type 3 rotation curve - rotational velocity (blue) and velocity dispersion (red).}
  \label{fig:type3_rot}
\end{minipage}
\end{figure*}

This follows a very similar evolution to type 1, except that the dark sphere, which starts with a radius of 87 kpc, collapses and relaxes to a steady configuration more quickly. The gas sphere falls into this deep gravitational well and forms a galaxy disc, with with $4.5 \times 10^{10}$ M$_{\odot}$ in the stellar component and $3.3 \times 10^{10}$ M$_{\odot}$ in the gas component, 4.6 Gyrs after the start of   the simulation. Again this would correspond to a redshift of $z = 1.06$ if we take the start to be at $z = 100$. $R_{1/2}$ is estimated to be between 3.7 and 10.6 kpc, while the radius of the gas disc is estimated to be 50 kpc (see Fig. \ref{fig:type2}). Again the rotational velocity is flat out to the outer radius of the gas disc, with a maximum of 225 km s$^{-1}$ (see Fig. \ref{fig:type2_rot}, which also shows the gas velocity dispersion).  In this case the average velocity dispersion outside $R_{max}$ is 37 km s$^{-1}$.

\subsection{Type 3}

For this simulation the dark matter and gas are two separated spheres each of radius 175 kpc, with their centres separated by 350 kpc, with the dark sphere centred at $(x,y)=(0,0)$, and the gas sphere at $(350,0)$. The dark sphere collapses as normal to a central relaxed halo, while the gas sphere falls into the potential well of the halo. The gas goes through a collapse and re-assemble phase in the centre of the halo, and eventually relaxes to form a galaxy disc, with $2.6 \times 10^{10}$ M$_{\odot}$ in the stellar component and $8.5 \times 10^{10}$ M$_{\odot}$ in the gas component, 5.6 Gyrs after the start of the simulation. Again this would correspond to a redshift of $z = 0.82$ if we assume the start is at $z = 100$. The value of $R_{1/2}$ is estimated to be between 3.12 and 12.4 kpc, while the radius of the gas disc is estimated to be 70 kpc (see Fig. \ref{fig:type3}). Again the rotational velocity is flat out to the outer radius of the gas disc, with a maximum of 246 km/sec (see Fig. \ref{fig:type3_rot}, which also shows the gas velocity dispersion). In this case the average velocity dispersion outside $R_{max}$ is 42 km s$^{-1}$, and the rotation axis is tilted by approximately $5^{\circ}$ from the z-axis.

\subsection{Type 4}
 
\begin{figure*}
\begin{minipage}{.5\textwidth}
  \raggedright
  \includegraphics[width=\linewidth]{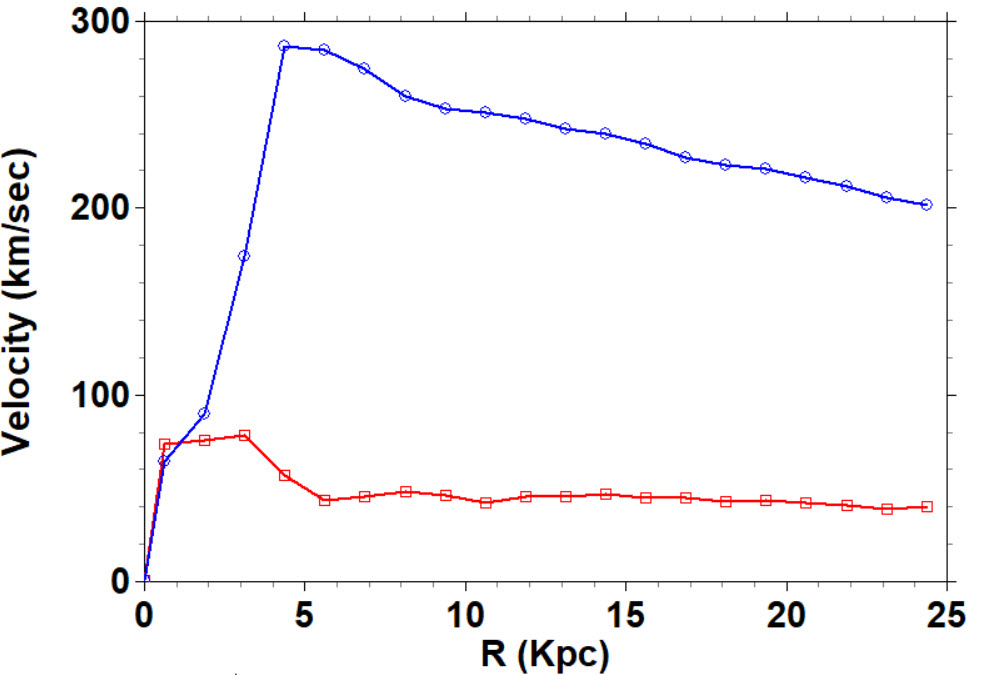}
  \caption{Type 4 rotation curve - rotational velocity (blue) and velocity dispersion (red).}
  \label{fig:type4_rot}
\end{minipage}
\begin{minipage}{.5\textwidth}
  \raggedleft
  \includegraphics[width=\linewidth]{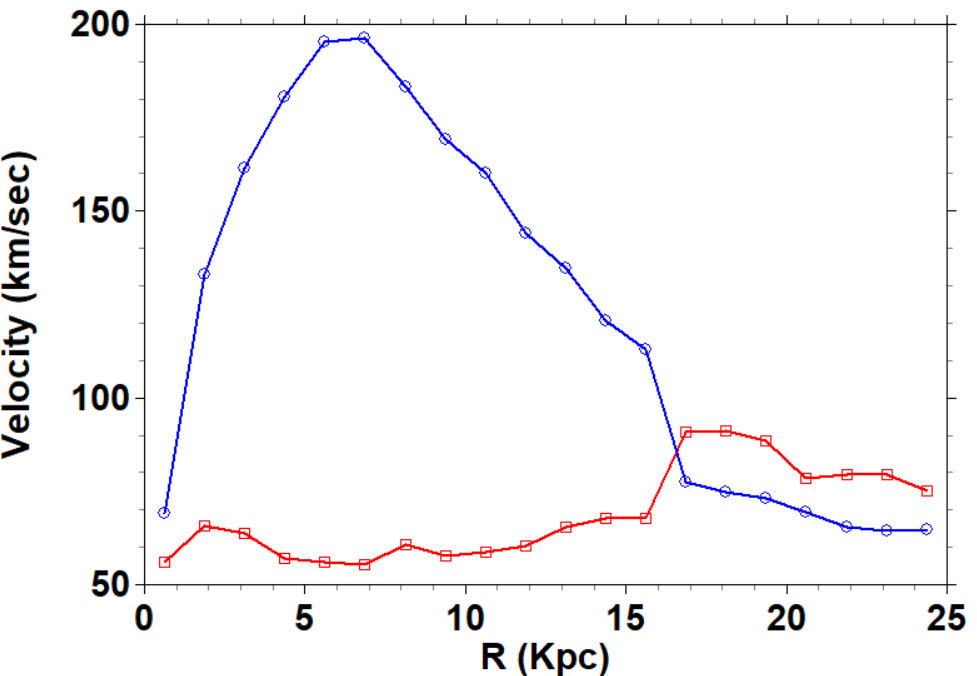}
  \caption{Type 5 rotation curve - rotational velocity (blue) and velocity dispersion (red).}
  \label{fig:type5_rot}
\end{minipage}
\end{figure*}

For this simulation the gas sphere is a uniform density sphere of radius 175 kpc with added noise, while the dark matter mass is split between six smaller spheres of radius 35 kpc, positioned at random within the gas sphere.  The dark spheres collapse as normal, and continue on orbits due to the initial rotational velocity about the centre of the gas sphere. At the same time the gas local to the dark spheres falls into their respective potential wells, and proceeds to form low mass disc galaxies. The dark halos then merge together, as well as their captured gas discs, forming a more massive halo/galaxy ensemble. The stellar mass of the galaxy disc is $4.3 \times 10^{10}$ M$_{\odot}$ and the gas mass is $3.4 \times 10^{10}$ M$_{\odot}$ at 5.3 Gyrs after the start of the simulation, that is at $z = 0.89$ if we assume the start is at $z = 100$.  The value of $R_{1/2}$ is estimated to be between 2.6 and 7.5 kpc, while the radius of the gas disc is estimated to be 80 kpc (see Fig. \ref{fig:type4}). The rotation axis of this disc galaxy is also no longer along the z-axis, but tilted by $15^{\circ}$ to it. After transforming particle positions and velocities, the rotation and dispersion curves of this object are shown in Fig. \ref{fig:type4_rot}. The maximum rotational velocity is 294 km s$^{-1}$, and the average velocity dispersion outside $R_{max}$ is 44 km s$^{-1}$.

\subsection{Type 5}

There may be other initial conditions which would lead to steeply declining rotation curves as in the Genzel result. However it is hard to see how a different configuration of dark matter in relation to the baryons would yield that, given the variety of configurations in types 1-4. The two most important physical quantities which have been found to affect the outcome of the simulations are the mass and angular momentum. Simulations were carried out with less angular momentum than that predicted by Peebles by a factor of 5, but these simply ended with smaller scale galaxies, but with equally high non-declining rotational velocities. The obvious quantity to vary would be the mass of the dark matter. Consequently simulations were carried out with no dark matter, and we present here an example of these. Type 5 has gas only initially with a mass of $8.5 \times 10^{10}$ M$_{\odot}$ in a sphere of radius 175 kpc, and an initial rotational velocity of 0.0644 Gyr$^{-1}$. It collapses under its own gravitational field to form a galaxy disc, with $2 \times 10^{10}$ M$_{\odot}$ in the stellar component and $5.9 \times 10^{10}$ M$_{\odot}$, in the gas component, 5.8 Gyrs after the start of the simulation. If we assume a standard $\Lambda$CDM cosmology this would correspond to a redshift of $z = 0.78$ if we take the start to be at $z = 100$. The value of $R_{1/2}$ is estimated to be between 3.7 and 16.8 kpc, while the radius of the gas disc is estimated to be 40 kpc (see Fig. \ref{fig:type5}). However in this case the rotation curve does decline steeply outside $R_{max}$. The rotation and dispersion curves of this object are shown in Fig. \ref{fig:type5_rot}, with maximum rotational velocity 195 km s$^{-1}$ and average velocity dispersion inside a 15 kpc radius of 59 km s$^{-1}$.

\subsection{Comparison with the Genzel et al. rotation curve}

To illustrate the contrast between these results and the rotation curves obtained by Genzel et al. (2017) using stacking, we normalise these rotation curves by $R_{max}$ and $V_{max}$ and plot the curves along with the data in Fig. 2b of Genzel et al. (rescaled in radius) in Fig. \ref{fig:comparison}.  

The rotation curves from types 1-4 are incompatible with the Genzel results. On the contrary type 5 can be seen to give a good match to the Genzel results.

\section{Discussion}

\begin{figure}[h]
\resizebox{\hsize}{!}{\includegraphics{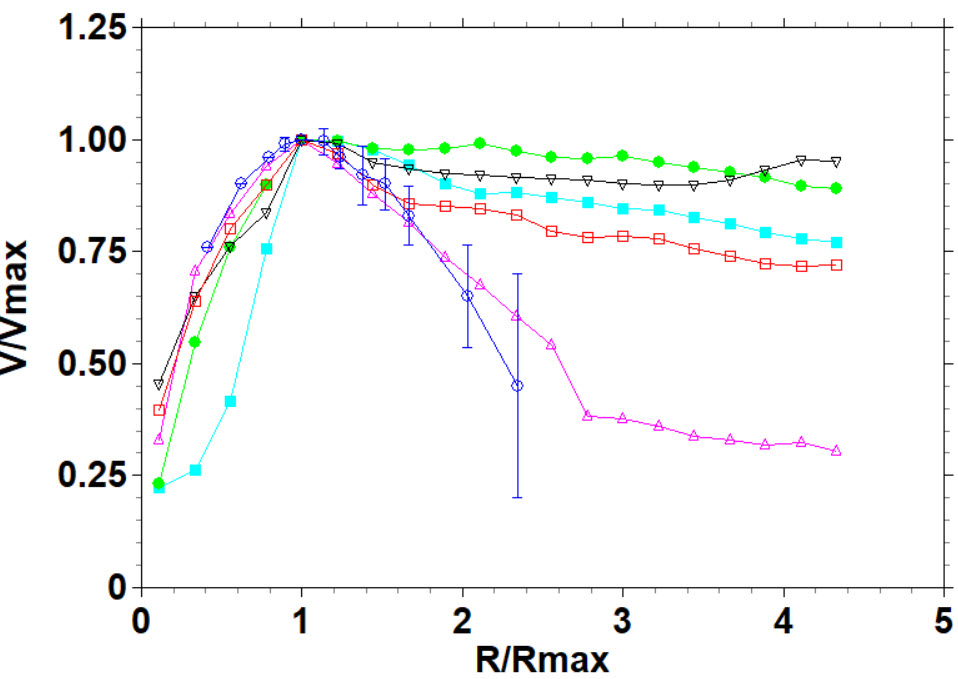}}
\caption{All rotational velocities versus radius, normalised to $R_{max}$ and $V_{max}$. Genzel et al. data (2017) - blue hollow circles with error bars, type 1 – red hollow squares, type 2 – black inverted triangles, type 3 – green filled circles, type 4 – turquoise filled squares, type 5 – mauve triangles}
\label{fig:comparison}
\end{figure}

These simulations demonstrate that if the formation scenario for galaxies is baryonic gas falling into the potential wells of formed, or forming, massive dark matter haloes, then as soon as a stellar disc is formed the gas rotation curve exhibits extended high velocity well beyond $R_{max}$. Fig. \ref{fig:comparison} clearly demonstrates this. The Genzel observations and the results of the simulation types 1-4 are obviously incompatible.

The type 1-4 rotation curves do decline slowly at large radius, but do not descend to half their maximum value until 6-17 $R_{max}$, corresponding to 62-74 kpc.  While the Genzel and type 5 rotation curves descend to half maximum in approximately 2 $R_{max}$, corresponding to 14 kpc in the case of type 5.

A possible explanation for the initial steeply falling rotation curves reported by Genzel et al. could be pressure support by turbulent motion in the interstellar medium. However that would require turbulent velocities $\sim 100$ km s$^{-1}$, and while, before corrections for beam smearing, the velocity dispersion detected by Genzel et al. has a value of that order at the centre of the galaxies, it quickly drops to $\sim 50$ km s$^{-1}$ outside $R_{max}$. Similarly low values for intrinsic velocity dispersion, that is $25-50$ km s$^{-1}$ for galaxies ranging from $z = 2.3-0.9$, were obtained by Wisnioski et al. (2015) - see also Ubler et al. (2019). This is in agreement with the simulations reported here, where the velocity dispersion of the gas has three contributions: random motion arising out of violent relaxation (typically 50-80 km/sec inside $R_{max}$,  but $\sim 30$ km s$^{-1}$ outside $R_{max}$, at formation of a stellar disc), turbulence from SN feedback (typically $\sim 30$ km s$^{-1}$ in the central peak but falling to $\sim 10$ km s$^{-1}$ outside $R_{max}$), and thermal motion (typically $\sim 10$ km s$^{-1}$ at all radii).

Teklu et al. (2018) have presented cosmological scale galaxy formation simulations in which a subset of galaxies exhibit declining rotation curves similar to the Genzel results within dark matter haloes, and argue that pressure support by velocity dispersion explains this. Indeed the plot of velocity dispersion which they display for the galaxies which have declining rotation curves has values $\sim 100$ km s$^{-1}$ at all radii, which implies that these simulated galaxies do have pressure support from velocity dispersion, unlike the Genzel et al. set. They report that 38\% of the galaxies they have studied in detail in their simulation have declining rotation curves, while the other 62\% have non-declining rotation curves. The 103 galaxies of the Genzel et al. set collectively exhibit falling rotation curves, which possibly implies that at high redshift falling rotation curves are universal. If that were the case then significant numbers of flat rotation curve galaxies from cosmological scale galaxy formation simulations such as those of Teklu et al. would be at odds with observations.

Finally, Ubler et al. (2021) have compared the observed kinematic results with the results of the IllustrisTNG cosmological scale simulations, and have concluded that the rotation curves from those simulations imply significantly more dark matter in the simulated galaxies than is indicated by the observations. This supports the conclusion here that the standard galaxy formation scenario within the $\Lambda$CDM cosmology, namely that the baryonic gas forming the gaseous and stellar discs collapses by falling into dark matter halos, may be at variance with at least a subset of observations of high redshift galaxy rotation.

\section{Conclusion}

The evidence to establish that steeply falling rotation curves are universal in galaxies with $z\gtrsim 1$ is not yet available. While the sample of galaxies used by Genzel et al. to stack rotation curves, thus extending the rotation curve beyond $R_{max}$, is large, actual measurement of rotational velocity in individual galaxies out to several times $R_{max}$, for a large sample, is required in order to justify such a statement. Recently Puglisi et al. (2023) published observations of 22 galaxies with $z \sim 1.5$, claiming that they have rotation curves flat out to $6 R_{d}$.  However only one of the observed rotation curves extends beyond $R_{max}$; the velocities at $6 R_{d}$ are obtained from a modelled extrapolation. To obtain actual measurements of velocity in the outer parts of the high redshift galaxies requires spectroscopy with higher sensitivity and resolution, such as may shortly be obtained using, for instance, the James Webb Space Telescope.

If such observations are obtained, confirming the results of Genzel et al. in a large sample of individual galaxies, then the simulations presented here cast serious doubt on the standard scenario that galaxies form by baryonic gas falling into the potential wells of massive dark matter halos, since the simulations reported here indicate that as soon as a stellar disc is formed, with velocity dispersion $\lesssim 50$ km s$^{-1}$, the extended rotation curve under these circumstances will not steeply decline.

The only variation in initial conditions that we could find to reproduce the Genzel results was to remove the dark matter halo in the initial conditions. If no alternative set of initial conditions including a dark halo can be found to reproduce the Genzel results, and they turned out be universal, then this would call into question the very existence of dark matter, since, due to the postulated non-dissipative nature of cold dark matter, it would seem unlikely that dark matter would be captured by a purely baryonic galaxy to produce a flat rotation curve at low $z$.

Another explanation of the flat rotation curves in contemporary galaxies would have to be found, possibly due the effect of galactic magnetic fields on gas rotation (Nelson 1988; Ruis-Granados et al. 2016). If the steeply falling rotation curves in the range $0.6< z <2.6$ reported by Genzel et al., and the flat rotation curves in the range $0.57< z <1.04$ reported by Sharma et al. are both confirmed, then one explanation could be that the roughly 2 Gyr interval between the median $z$ of the two respective galaxy sets could be sufficient for the galaxy magnetic field to be generated to a sufficient magnitude to be dynamically significant. The growth time for the magnetic field in a differentially rotating turbulent gas disc is approximately 1-2 Gyrs (Panesar \& Nelson 1992).

On the other hand there remains the possibility that some galaxies do form in dark matter haloes, while others do not. The properties of dark matter and baryonic matter are sufficiently different that no satisfactory explanations has so far been put forward as to why they should be locked together in the turbulent primordial universe (Nelson 2022).

\begin{appendix}

\section{Fits to the NFW profile}
\label{app:nfw}

\begin{figure*}[ht!]
\centering
\includegraphics[width=16cm]{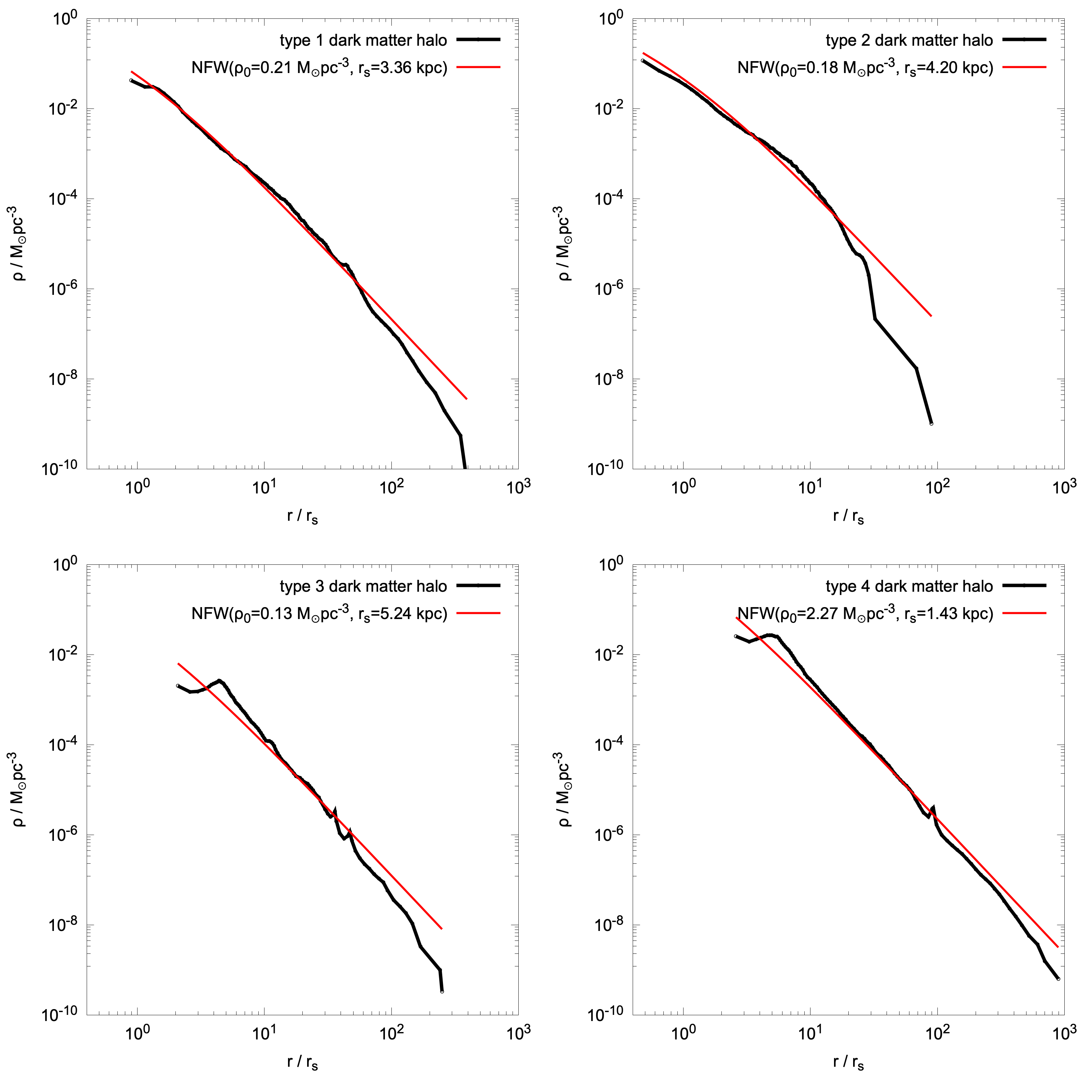}
\caption{Simulation halo density profile $\rho(r)$ plotted in black with the corresponding $\rho_{\text{NFW}}(r)$ in red for each of types 1-4.}
\label{fig:nfw}
\end{figure*}

Since the potential well of the dark matter haloes in simulation types 1-4 is crucial in determining the rotation curves of these galaxies, it is relevant to examine the nature of the haloes produced in the simulations. Consequently here, in addition to the images show in Appendix \ref{app:images} and the associated videos, we plot volume mass density radial profiles of the dark matter, and fit these to the canonical NFW profile (Navarro, Frenk \& White 1996).

The NFW profile as a function of radius r is
\begin{equation}
\rho_{\text{NFW}}(r) = \frac{\rho_{0}}{\zeta {(1 + \zeta )}^{2}}
\end{equation}
where $\rho_{0}$ is the characteristic density, $\zeta$ is $r/r_{s}$, and $r_{s}$ is the scale radius. Integrating this over the volume of a spherical halo, we obtain for $M_{\text{NFW}}(r)$, the total
mass inside radius $r$,
\begin{equation}
M_{\text{NFW}}(r) = 4\pi \rho_{0} r_{s}^{3} \left[ \log (1+\zeta) - \frac{\zeta}{1+\zeta} \right]
\label{eq:nfw}
\end{equation}

Dark matter particles were binned radially in shells such that there were 1000 particles per bin, and the particle masses in each shell summed. The mass profile $M(r)$ was produced by summing shells cumulatively with increasing radius. The NFW mass profile in equation \ref{eq:nfw} was then fit to $M(r)$ using a non-linear least-squares algorithm, and $\rho_{0}$ and $r_{s}$ as fitting parameters. By dividing the mass in each shell by the volume of the shell at $r$, $\rho(r)$ was computed for the simulation halo. This was plotted along with $\rho_{\text{NFW}}(r)$ in Fig. \ref{fig:nfw}.

Over most of the radial range there is a remarkably good agreement with the NFW profile in spite of the oblate spheroidal shape of the simulated haloes, and the highly non spherical initial distribution of the dark matter in type 4. This demonstrates the effectiveness of the violent relaxation process in settling the component into a stable configuration.

The only  significant deviation from the NFW profile occurs at radii $>100$ kpc, well outside the galaxy radii, and where the low density of dark material would diminish the integrity of the simulation of the dark component.

\section{Estimation of Obscuration}
\label{app:obsc}

The mass of a star particle in the simulations is $2.5 \times 10^{6}$ M$_{\odot}$, therefore they represent multiple stars, hence the surface luminosity of the stellar component in the simulations can be estimated by simply summing the surface density of star particles. However the simulated galaxies have evolved to the point that significant dust will have been ejected into the interstellar medium, sufficient to cause obscuration varying with radius from the centre of the galaxy.

The generation of dust and its obscuring effects was the subject of Churches, Nelson \& Edmunds (2001, 2004), who estimated the generation of heavy elements and the obscuration by dust in the context of simulations of galaxy formation using the same code as in this paper. The optical depth of the galaxy disc is given by equation (7) of Churches, Nelson \& Edmunds (2004)
\begin{equation}
\tau \approx 4 \times 10^{-24} \text{Z} \text{N}_{g}
\label{eq:tau}
\end{equation}
where Z is the metallicity, and N$_{g}$ is the surface number density of gas in m$^{-2}$. Fig. 2a of the 2001 paper gives the quantity 12+log(O/H) as a function of radius and time, where O/H is the abundance of Oxygen relative to hydrogen. The sloping lines in that figure can be approximated by straight lines with the following equations
\begin{equation}
\text{12+log(O/H)} = 9.1 - 0.05 r \text{ where } t_{s}=5 \text{ Gyr},
\label{eq:oh1}
\end{equation}
and
\begin{equation}
\text{12+log(O/H)} = 8.6 - 0.042 r \text{ where } t_{s}=2 \text{ Gyr},
\label{eq:oh2}
\end{equation}
where $t_{s}$ is the time since the start of the simulation.

The values for intermediate times can be obtained by interpolation or extrapolation of the straight line coefficients. O/H is related to Z by the approximate equation (Pagel et al., 1992)
\begin{equation}
\text{Z} \sim 23\text{(O/H)}.
\label{eq:z}
\end{equation}
By evaluating the gas surface density in the final frames of each simulation type, equations \ref{eq:tau}, \ref{eq:oh1}, \ref{eq:oh2}, and \ref{eq:z} were used to evaluate the optical depth of the dust as a function of radius.

However the simulations do not have sufficient resolution in the vertical direction to properly resolve the expected difference of scale height in the gas and stars -- the two components have similar rms heights from the simulation. Consequently to calculate the amount of obscuration as a function of radius we use the two extreme cases of (1) dust concentrated in a central layer much smaller then the stellar disc thickness, and (2) dust distributed throughout the stellar disc thickness.

In case 1 the reduction factor of luminosity of the stars by the dust ($RF$) given optical depth $\tau$, is given by
\begin{equation}
RF = 0.5(1 + \exp -\tau)
\end{equation}
while in case 2
\begin{equation}
RF = (1 - \exp -\tau)/\tau.
\end{equation}
These two cases were applied to the luminosity of the stellar component represented by the surface density of stellar particles, and the resulting distribution was then used to estimate $R_{1/2}$. The consequent values of $R_{1/2}$ range from 2.6 to 4.2 kpc (average 3.1) over the 5 types in case 1, and 7.5 to 16.8 kpc (average 11.8) in case 2 , whereas without taking the dust obscuration into account the range is 1.8  to 3.7 kpc (average 3.0), due to the higher unobscured luminosity at the centre.

\section{Galaxy Images}
\label{app:images}

\begin{figure*}[ht]
\centering
\includegraphics[width=16cm]{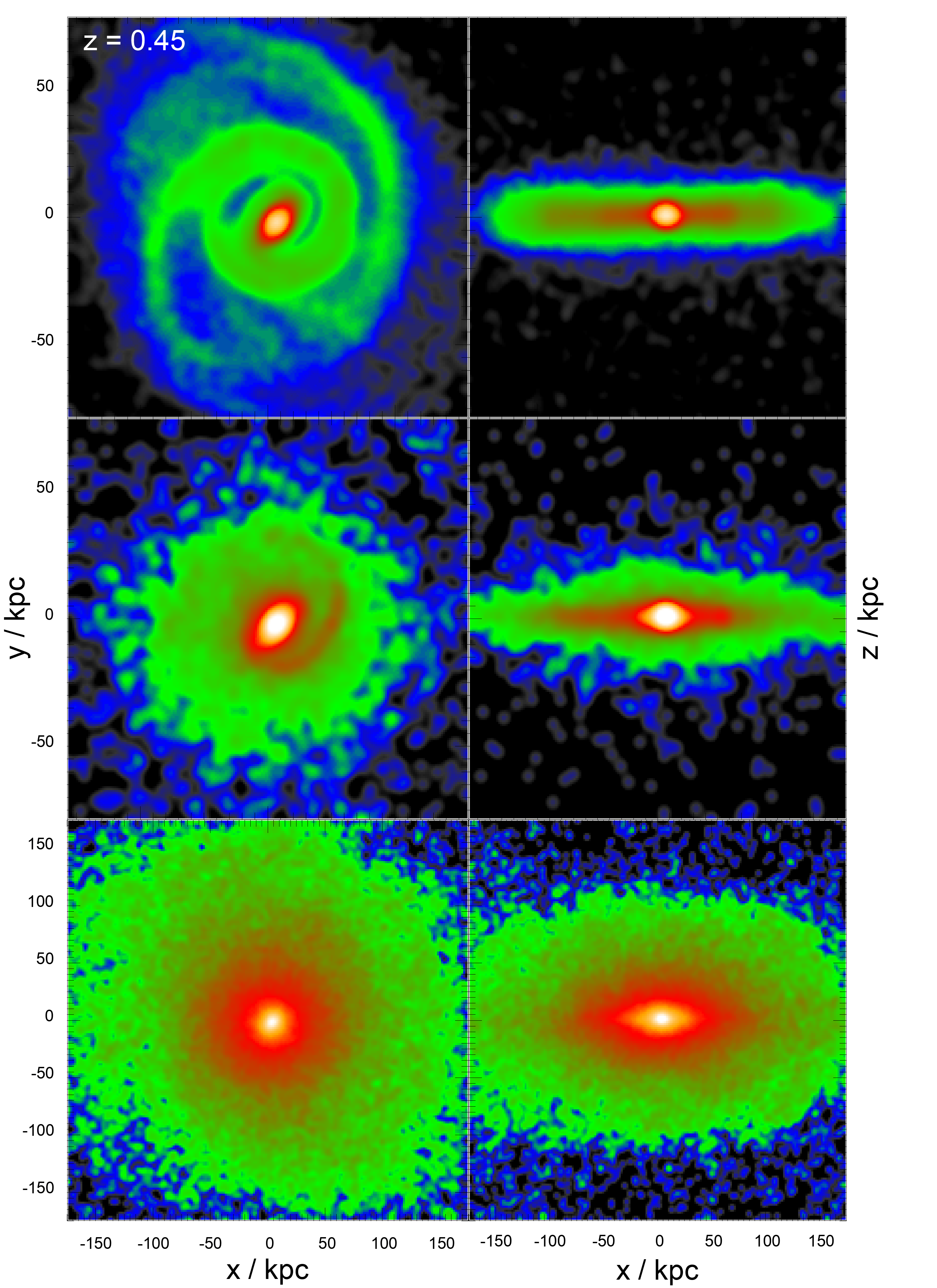}
\caption{Type 1 images at $z = 0.45$. Top to bottom - gas, stars, dark matter; left to right - face-on, edge-on. The side of the panels is 157.5 kpc for gas and stars showing the central regions in more detail, and 350 kpc for dark matter. For an animation of this run showing the particle images on YouTube use this link \href{https://youtu.be/bRrs5Hu6jYo}{\textcolor{blue}{\textbf{Type 1}}}. The panels in the animation are in the same configuration as this figure, but all with width 350 kpc. Density contours in this figure are rendered logarithmically, from black ($< 0.05$  M$_{\odot}$pc$^{-2}$), blue ($\sim 0.5$  M$_{\odot}$pc$^{-2}$), green ($\sim 5$ M$_{\odot}$pc$^{-2}$), red ($\sim 50$ M$_{\odot}$pc$^{-2}$), to white ($>1000$  M$_{\odot}$pc$^{-2}$).}
\label{fig:type1}
\end{figure*}

Figs. \ref{fig:type1} to \  \ref{fig:type5} show images of the surface density for gas, stars, and dark matter at the time in the simulations where the rotation curves were measured, for simulation types 1-5.

\begin{figure*}[ht]
\centering
\includegraphics[width=16cm]{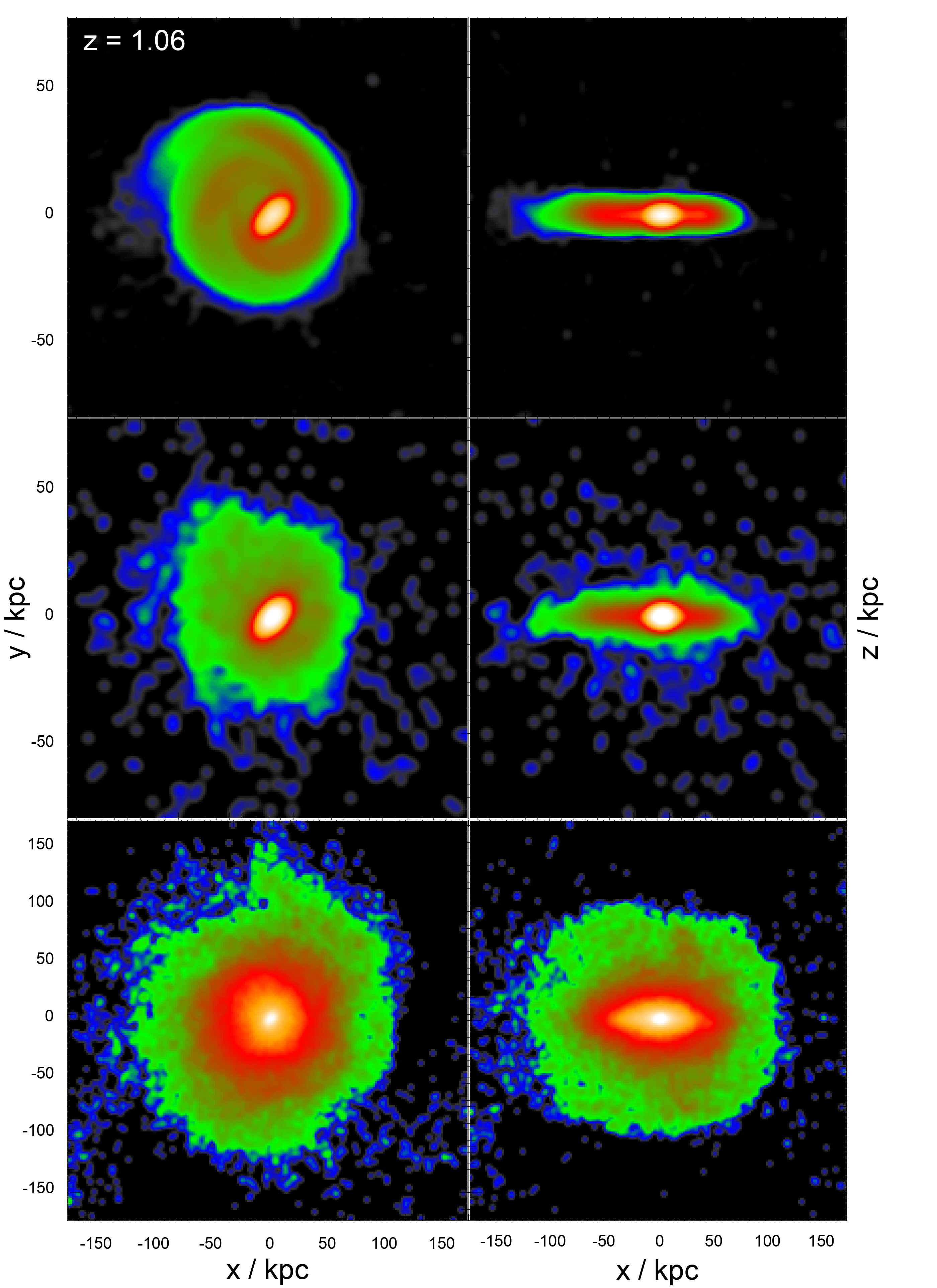}
\caption{Type 2 images at $z = 1.06$. Top to bottom - gas, stars, dark matter; left to right - face-on, edge-on. The side of the panels is 157.5 kpc for gas and stars showing the central regions in more detail, and 350 kpc for dark matter. For an animation of this run showing the particle images on YouTube use this link \href{https://youtu.be/GGSY-yLBqcg}{\textcolor{blue}{\textbf{Type 2}}}. The panels in the animation are in the same configuration as this figure, but all with width 157.5 kpc. Density contours in this figure are rendered logarithmically, from black ($< 0.05$  M$_{\odot}$pc$^{-2}$), blue ($\sim 0.5$  M$_{\odot}$pc$^{-2}$), green ($\sim 5$ M$_{\odot}$pc$^{-2}$), red ($\sim 50$ M$_{\odot}$pc$^{-2}$), to white ($>1000$  M$_{\odot}$pc$^{-2}$).}
\label{fig:type2}
\end{figure*}

\begin{figure*}[ht]
\centering
\includegraphics[width=16cm]{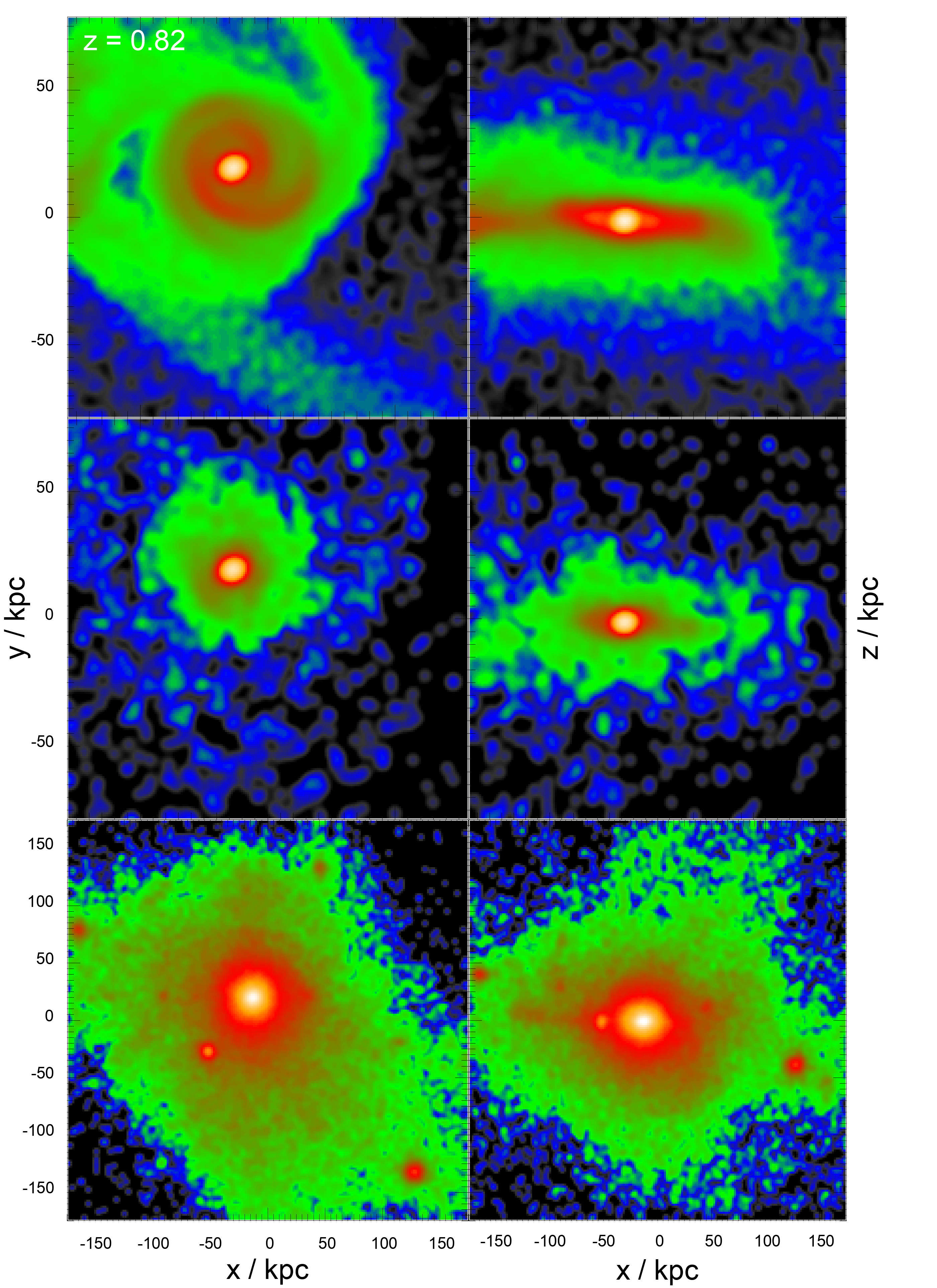}
\caption{Type 3 images at $z = 0.82$. Top to bottom - gas, stars, dark matter; left to right - face-on, edge-on. The side of the panels is 157.5 kpc for gas and stars showing the central regions in more detail, and 350 kpc for dark matter. For an animation of this run showing the particle images on YouTube use this link \href{https://youtu.be/mtwIU7Qh_g8}{\textcolor{blue}{\textbf{Type 3}}}. The panels in the animation are in the same configuration as this figure, but all with width 350 kpc. Density contours in this figure are rendered logarithmically, from black ($< 0.05$  M$_{\odot}$pc$^{-2}$), blue ($\sim 0.5$  M$_{\odot}$pc$^{-2}$), green ($\sim 5$ M$_{\odot}$pc$^{-2}$), red ($\sim 50$ M$_{\odot}$pc$^{-2}$), to white ($>1000$  M$_{\odot}$pc$^{-2}$).}
\label{fig:type3}
\end{figure*}

\begin{figure*}[ht]
\centering
\includegraphics[width=16cm]{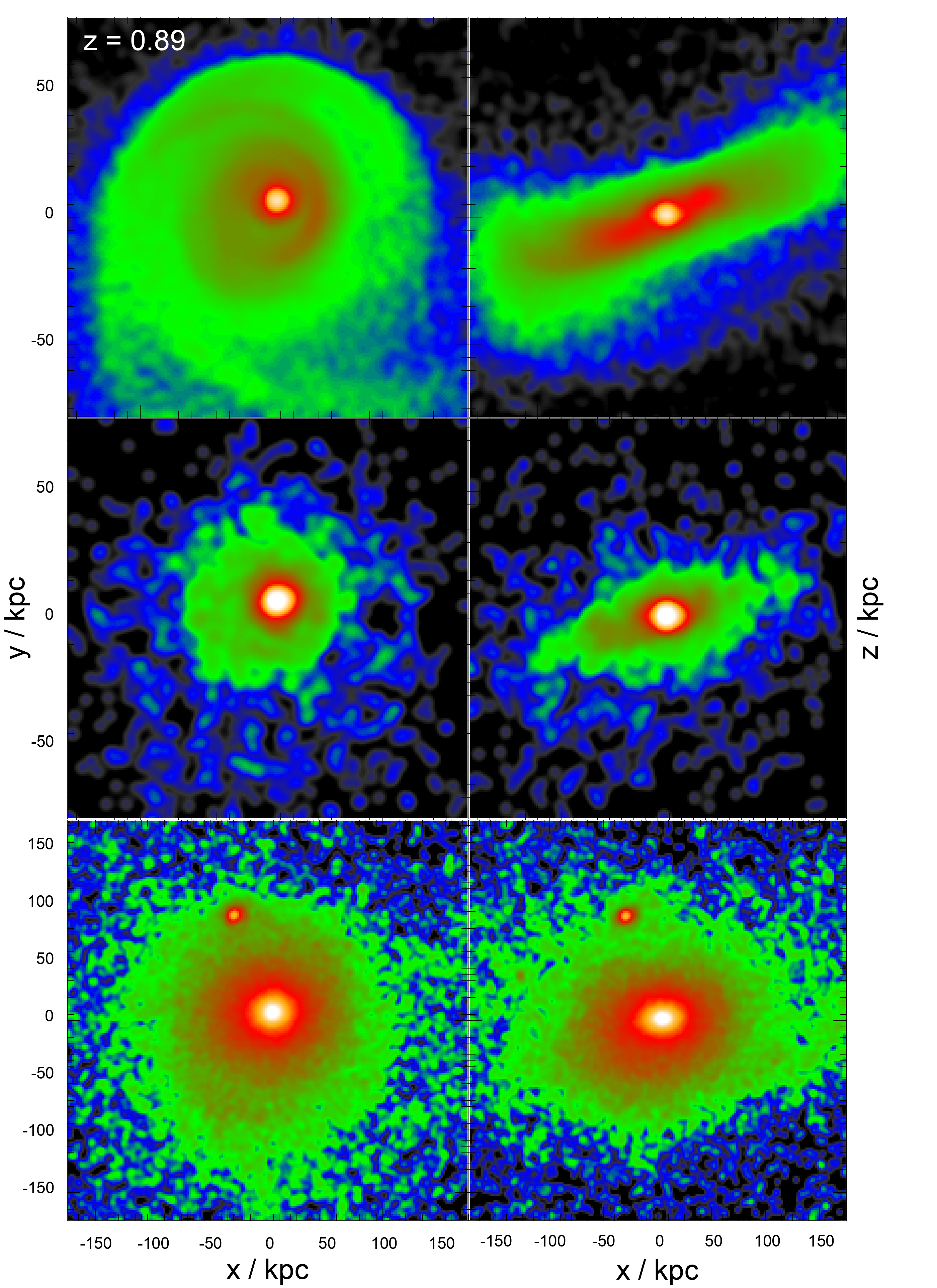}
\caption{Type 4 images at $z = 0.89$. Top to bottom - gas, stars, dark matter; left to right - face-on, edge-on. The side of the panels is 157.5 kpc for gas and stars showing the central regions in more detail, and 350 kpc for dark matter. For an animation of this run showing the particle images on YouTube use this link \href{https://youtu.be/3j2rUXx7NOQ}{\textcolor{blue}{\textbf{Type 4}}}. The panels in the animation are in the same configuration as this figure, but all with width 350 kpc. Density contours in this figure are rendered logarithmically, from black ($< 0.05$  M$_{\odot}$pc$^{-2}$), blue ($\sim 0.5$  M$_{\odot}$pc$^{-2}$), green ($\sim 5$ M$_{\odot}$pc$^{-2}$), red ($\sim 50$ M$_{\odot}$pc$^{-2}$), to white ($>1000$  M$_{\odot}$pc$^{-2}$).}
\label{fig:type4}
\end{figure*}

\begin{figure*}[ht]
\centering
\includegraphics[width=16cm]{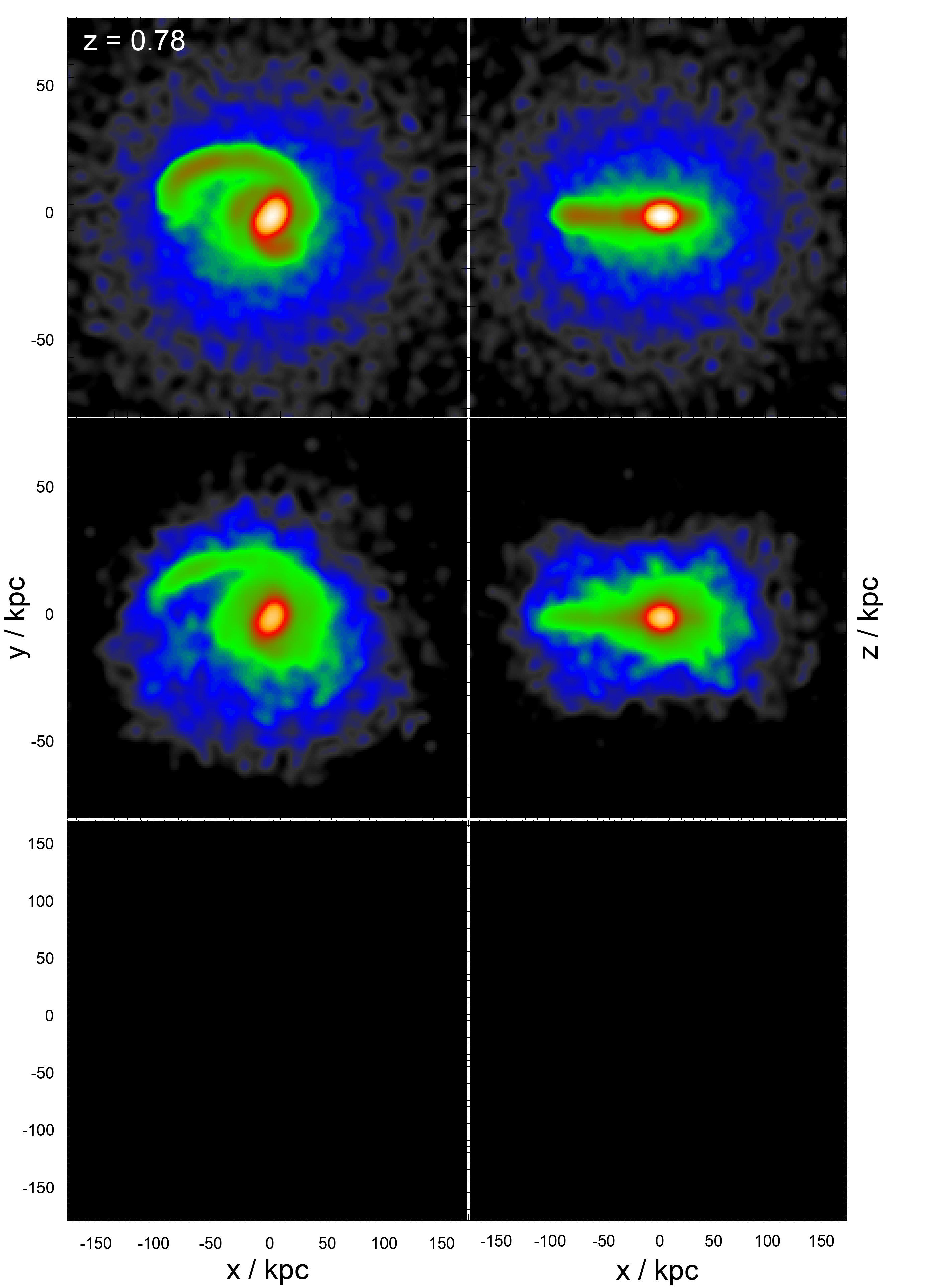}
\caption{Type 5 images at $z = 0.78$. Top to bottom - gas, stars; left to right - face-on, edge-on. The side of the panels is 157.5 kpc for gas and stars showing the central regions in more detail. (Note that there was no dark matter component included in this simulation.) For an animation of this run showing the particle images on YouTube use this link \href{https://youtu.be/NCnqIOLRyDg}{\textcolor{blue}{\textbf{Type 5}}}. The panels in the animation are in the same configuration as this figure, and also have a width of 157.5 kpc. Density contours in this figure are rendered logarithmically, from black ($< 0.05$  M$_{\odot}$pc$^{-2}$), blue ($\sim 0.5$  M$_{\odot}$pc$^{-2}$), green ($\sim 5$ M$_{\odot}$pc$^{-2}$), red ($\sim 50$ M$_{\odot}$pc$^{-2}$), to white ($>1000$  M$_{\odot}$pc$^{-2}$).}
\label{fig:type5}
\end{figure*}

\end{appendix}
\end{document}